\documentclass{jnst} 
\usepackage{comment}
\usepackage{url}
\usepackage[tablesfirst,notablist]{endfloat} 
\usepackage{color}
\def\newblock{\hskip .11em plus .33em minus .07em}

\title{$^{235}$U(n, f) Independent Fission Product Yield and Isomeric Ratio Calculated with the Statistical Hauser-Feshbach Theory}

\author{Shin Okumura$^{1}$\thanks{Corresponding author. Email: okumura.s.aa@m.titech.ac.jp}, Toshihiko Kawano$^{2}$, Patrick Talou$^{2}$, Patrick Jaffke$^{2}$, and Satoshi Chiba$^{1}$}

\inst{$^{1}$Laboratory for Advanced Nuclear Energy, Tokyo Institute of Technology, 2-12-1 Ookayama, Meguro-ku, Tokyo 152-8550, Japan.;
$^{2}$Theoretical Division, Los Alamos National Laboratory, Los Alamos, NM 87545, USA}

\abst{
We have developed a Hauser-Feshbach fission fragment decay model,
HF$^3$D, which can be applied to the statistical decay of more than
500 primary fission fragment pairs (1,000 nuclides) produced by the
neutron induced fission of $^{235}$U.  The fission fragment yield
$Y(A)$ and the total kinetic energy TKE are model inputs, and we
estimate them from available experimental data for the
$^{235}$U(n$\rm_{th}$,f) system. The model parameters in the
statistical decay calculation are adjusted to reproduce some fission
observables, such as the neutron emission multiplicity
$\overline{\nu}$, its distribution $P(\nu)$, and the mass dependence
$\overline\nu(A)$.  The calculated fission product yield and isomeric
ratio are compared with experimental data.  We show that the
calculated independent fission product yield $Y_I(A)$ at the thermal
energy reproduces the experimental data well, while the calculated
isomeric ratios tend to be lower than the Madland-England model
prediction. The model is extended to higher incident neutron energies
up to the second chance fission threshold.  We demonstrate for the
first time that most of the isomeric ratios stay constant, although
the production of isomeric state itself changes as the incident energy
increases.}
\kword{Nuclear Fission, Statistical Decay, Isomeric Ratio, Fission Product Yield}
\begin{document}

\maketitle
\section{Introduction}
\label{sec:Introduction}
Thermal neutron induced fission, such as the $^{235}$U(n$_{\rm th}$,f)
  or $^{239}$Pu(n$_{\rm th}$,f) reactions, produces roughly 800
  primary fission fragments \cite{LA-UR-94-3106}. Since these fission
  fragments are highly excited, they de-excite by emitting several
  prompt neutrons and $\gamma$ rays to reach their ground or
  metastable states within a time-scale of compound nucleus in the
  fission process. The independent fission product yield $Y_I(Z, A)$,
  which is a distribution of nuclides after emission of the prompt
  particles but before beta decay, plays an important role in many
  applications such as estimation of decay heat \cite{Yoshida1981,
  Yoshida2008, ChibaGo2017} and delayed neutron emission
\cite{Yoshida2002, ChibaGo2013} in nuclear reactors, the reactor
neutrino study \cite{Hayes2015}, prediction of fission product
inventory at each stage of the nuclear fuel cycle, the radio-isotope
production for medical applications, development of advanced reactor
and transmutation systems, fission in the galactic chemical evolution
\cite{Shibagaki2016}, and so on. A demand for high quality data of
fission product yield (FPY) in such applications is rapidly
increasing. New applications may require accurate FPY data at several
neutron incident energies, while the current evaluated FPY data files
contain only three energy points; the thermal, fast and 14-MeV
incident energies, with an exception of the $^{239}$Pu file in the
evaluated nuclear data file, ENDF/B-VII.1 \cite{Chadwick2010}, where
two energy points (0.5 and 2~MeV) are given in the fast
range\cite{Kawano2013b}.

Significant efforts have been made to compile the FPY experimental data
and evaluation for the nuclear data libraries in the past. England and
Rider \cite{LA-UR-94-3106} evaluated the FPY data for 60 fissioning
systems in 1994, and released their results as part of
ENDF/B-VI. They calculated $Y_I(Z, A)$ by combining the evaluated mass
chain yield and the charge distribution in the most probable charge
($Z_p$) model proposed by Wahl \cite{LA13928}. The FPY evaluation in
Japanese Evaluated Nuclear Data Library
(JENDL/FPY-2011) \cite{JAEA-DATACODE-2011-025, Katakura2016} followed
the procedure of England and Rider, and these FPY data were upgraded
by including new experimental FPY and making it consistent with the
updated JENDL decay data library (JENDL/FPD-2011)
\cite{JAEA-DATACODE-2011-025}.

The isomeric ratio data in both libraries are estimated by
introducing the Madland and England (ME) model \cite{Madland1977,
  LA-6595-MS}, whenever the isomeric ratio is experimentally
unknown. Relative population of the isomeric and ground states in a
fission product (after prompt neutron emission) is calculated by
looking at the difference in their spins.  An even-odd effect in the
fission products is also considered. While this model is widely used
in the current libraries, recent advances in the fission model
together with the statistical Hauser-Feshbach decay \cite{Kawano2013,
  Stetcu2013} could also be able to improve the evaluation of FPY
data.

Despite many theoretical studies on fission that have
been made in the past, prediction and reproduction of all the fission
observables in a consistent manner are still challenging. A modeling of
de-excitation of fission fragments requires many physical
quantities \cite{Becker2013} that define an initial configuration of
the fragment decay, such as the fragment excitation energy,
spin-parity distribution, and $(Z, A)$ distribution of primary fission
fragments. The model predicts prompt particle emission probabilities
and multiplicities, and $Y_I(Z, A)$ simultaneously by integrating over
the distribution of initial configurations. Instead of performing the
integration, several Monte Carlo (MC) tools have been developed to
calculate the fission de-excitation process and to reproduce these
observables
\cite{Becker2013,Litaize2010, Randrup2009, Randrup2014, Schmidt2016, Talou2017}.
Although the MC technique gives correlations in the prompt particle
emissions in fission, and it facilitates experimental data analysis by
emulating directly the measurement set-up, its lengthy computation
makes it difficult to fine-tune their model parameters. In addition,
when a probability of fission fragment production is extremely small,
the MC technique never samples such a case in a reasonable
computational time. Because FPY varies in the order of magnitude, {\it
e.g.} typically from 10$^{-12}$ to 10$^{-2}$ given in the evaluated
files, the MC technique is not an efficient method to be adopted in
the FPY evaluation, and one has to migrate to a deterministic method,
in which all the possible fission fragments can be included.

Our approach is to develop a new method to evaluate $Y_I(Z, A)$ by
applying the deterministic technique for the primary fission fragment
decay. The outline of our method is similar to our past study
\cite{Kawano2013}, albeit no MC sampling is performed anymore. We
apply the Hauser-Feshbach statistical decay to about 500 primary
fission fragment pairs (1000 nuclides), and abbreviate this as HF$^3$D
(Hauser-Feshbach Fission Fragment Decay). Since the spin and parity
conservation is naturally involved in the HF$^3$D model, the isomeric
ratio calculation is just straightforward \cite{Stetcu2013}. The mass
distribution of the primary fission fragment is represented by five
Gaussians, and the Gaussian parameters are fitted to experimental data
of $^{235}$U(n$\rm_{th}$,f). We use Wahl's $Z_p$ model for the charge
distribution. The experimental data of average prompt neutron
multiplicity $\overline{\nu}$, its distribution $P(\nu)$, the mass
dependence $\overline{\nu}(A)$, and the total kinetic energy (TKE) are
taken into account to constrain our model parameters. A dynamical
treatment of fission process such as the Langevin model
\cite{Aritomo2013, Aritomo2014, Usang2016, Ishizuka2017, Usang2017, Sierk2017} or a random walk technique
\cite{Randrup2011a, Randrup2011b, Moller2015} is able to provide some
of our inputs such as TKE and/or $Y(Z,A)$ distribution. These
dynamical models could be used when their predictive capability meets
required accuracy of the evaluated FPY data. At this moment we stay on
a phenomenological approach and rely more on the available
experimental data for practical purposes. Once the model parameters in
HF$^3$D are fixed to the thermal neutron induced fission data, we
extrapolate our calculation up to the threshold energy of the second
chance fission. The energy dependence in FPY and isomeric ratio is
thus calculated.

\section{Modeling for Statistical decay of fission fragments}
\label{sec:Modeling}
\subsection{Generating fission fragment pairs}

In the HF$^3$D model, we apply the statistical Hauser-Feshbach theory
to the fission fragment decay process, in which a competition between
the neutron and $\gamma$-ray channels is properly included at all the
compound decay stages.  Instead of employing the MC technique to the
Hauser-Feshbach theory \cite{Litaize2010, Becker2013}, HF$^3$D
numerically integrates the emission probabilities over the
distribution of fission fragment yield, as well as the distributions
of excitation energy, spin and parity in each fragment. Although
this deterministic method loses information on the correlated particle
emission \cite{Talou2017}, it allows us to include a lot of fission
fragment pairs that could have a tiny fission yield and never been sampled
by the MC technique. This is particularly important for studying the
production of radioactive isotopes \cite{Stetcu2013}.

We assume that the fission fragment mass distribution $Y(A)$ is
approximated by five Gaussians,
\begin{equation}
 Y(A) = \sum_{i=1}^5 \frac{F_i}{\sqrt{2\pi}\sigma_i}
        \exp\left\{
              -\frac{(A-A_m + \Delta_i)^2}{2\sigma_i^2}
            \right\} \ ,
\label{eq:fivegaussian}
\end{equation}
where $\sigma_i$ and $\Delta_i$ are the Gaussian parameters, the index
$i$ runs from the low mass side, and the component of $i=3$ is for the
symmetric distribution ($\Delta_3 = 0$). $A_m=A_{\rm CN}/2$ is the
mid-point of the mass distribution, $A_{\rm CN}$ is the mass number of
fissioning compound nucleus, and $F_i$ is the fraction of each
Gaussian component.

For the charge distribution $Z(A)$ we apply Wahl's $Z_p$ model
\cite{Wahl1988} with the parameters given in
Ref~\cite{LA13928}. Combining the mass and charge distributions, we
obtain the primary fission fragment distribution $Y(Z,A)$. Because the
distribution is symmetric with respect to $A_m$,
\begin{equation}
  Y(Z_l,A_l) = Y(Z_{\rm CN}-Z_l,A_{\rm CN}-A_l) = Y(Z_h,A_h) \ ,
\end{equation}
where $l$, $h$ and CN denote the light, heavy fragment and
compound nucleus, respectively. We abbreviate this by $Y_k$, where $k$
stands for the $k$-th fission fragment pair of
$(Z_l,A_l)$-$(Z_h,A_h)$.

Setting the lowest mass number $A_{\rm min}$ to 50, $A_{\rm max} =
A_{\rm CN} - A_{\rm min} = 186$ for the n + $^{235}$U case, there will
be more than 750 pairs of light and heavy fission fragments, namely
more than 1,500 fission fragments. The energy conservation --- the sum
of the total excitation energy TXE and the total kinetic energy TKE
cannot exceed the reaction $Q$-value --- reduces the number of
possible pairs. We fit a simple analytic function
\begin{eqnarray}
  {\rm TKE}(A_h)
  &=& (p_1 - p_2 A_h)
      \left\{
        1 - p_3 \exp\left(-\frac{(A_h - A_m)^2}{p_4}\right)
      \right\} \nonumber\\
  &+& \epsilon_{\rm TKE} \ ,
  \label{eq:TKE}
\end{eqnarray}
to available experimental data of TKE at thermal energy, where $p_1$
-- $p_4$ are the fitting parameters, and a small correction of
$\epsilon_{\rm TKE}$ ensures the average of Eq.~(\ref{eq:TKE}) agrees
with the evaluated $\overline{\rm TKE}$.  Since ${\rm TKE}(A)$ is
already averaged over the charge distribution $Z(A)$ for a fixed $A$
number, we assume that the TKE of fragment pairs having the same $A$
distributes according to their charge product $Z_l Z_h$, such that the
average of different $Z$'s coincides with Eq.~(\ref{eq:TKE}).  TKE for
a given $(Z_l,A_l)$-$(Z_h,A_h)$ pair is denoted by
TKE$(Z_l,A_l,Z_h,A_h)$, and TXE is calculated as
\begin{align}
  {\rm TXE}(Z_l,A_l,Z_h,A_h)
  &= E_{inc} + B_n + [M_n(Z_{CN},A_{CN}) -  M_n(Z_l,A_l) - M_n(Z_h,A_h)]c^2\nonumber\\
  &\qquad\quad- {\rm TKE}(Z_l,A_l,Z_h,A_h) \ ,
  \label{eq:TXE}
\end{align}
where $E_{inc}$ is the incident neutron energy in the
center-of-mass system, and $B_n$ is the neutron binding energy of the
target, and $M_n$ represents the nuclear mass. When TXE becomes
negative, we eliminate such pairs.

\subsection{Hauser-Feshbach approach to the fission fragment decay}

In the deterministic method, physical quantities that can be compared
with experimental data are given by a fragment-yield weighted sum of
the calculated results. For example, the average number of prompt 
neutrons $\overline{\nu}$ is calculated as
\begin{equation}
  \overline{\nu}
   = \sum_{k=1}^N Y_k
     \left(
        \overline{\nu}_l^{(k)} + \overline{\nu}_h^{(k)}
     \right) \ ,
\end{equation}
where $N$ is the total number of fragment pairs, and the neutron
multiplicities $\overline{\nu}_{l,h}^{(k)}$ are given by integrating
the neutron evaporation spectrum $\phi_{l,h}^{(k)}$ from the
light or heavy fragment in the center-of-mass system,
\begin{equation}
  \overline{\nu}_{l,h}^{(k)}
   = \int\! dE_x \sum_{J\Pi} \int\! d\epsilon
     \ R(J,\Pi) G(E_x) \phi_{l,h}^{(k)}(J,\Pi,E_x,\epsilon) \ ,
\end{equation}
where $R(J,\Pi)$ is the probability of nucleus having the state of
spin $J$ and parity $\Pi$, and $G(E_x)$ is the distribution of
excitation energy. They satisfy the normalization condition of
$\sum_{J\Pi} R(J,\Pi) = 1$ and $\int G(E_x)dE_x = 1$.

For the spin and parity distributions we follow our previous work
\cite{Kawano2013}, in which the spin-parity population distribution is
expressed by
\begin{equation}
  R(J,\Pi) = \frac{J+1/2}{2f^2\sigma^2(U)}
             \exp\left\{-\frac{(J+1/2)^2}{2 f^2\sigma^2(U)} \right\} \ ,
  \label{eq:rjp}
\end{equation}
where the parity distribution is just 1/2, $\sigma^2(U)$ is the spin
cut-off parameter, $U$ is the excitation energy corrected by the
pairing energy $\Delta$ as $U=E_x-\Delta$, and $f$ is a scaling factor
determined later by comparing the calculated results with experimental
data.

We estimate the average excitation energies in each fragment, $E_l$ and
$E_h$, with the anisothermal model \cite{Ohsawa1999, Ohsawa2000,
  Kawano2013} that is characterized by an anisothermal parameter $R_T$
defined as the ratio of effective temperatures in the fission
fragments
\begin{equation}
   R_T = \frac{T_l}{T_h} = \sqrt{\frac{U_l}{U_h} \frac{a_h(U_h)}{a_l(U_l)}} \ ,
   \label{eq:rt}
\end{equation}
where $a(U)$ is the shell-effect corrected level density parameter at
the excitation energy of $U$. In reality, TKE in Eq.~(\ref{eq:TKE})
could have a distribution characterized by the width $\delta_{\rm
  TKE}$, which is empirically known to be about
8--10~MeV \cite{Baba1997,Hambsch}. This width propagates to the width
of TXE through $\delta_{\rm TXE} = \delta_{\rm TKE}$, then perturbs
the excitation energies of two fragments. Assuming Gaussian for the
TXE distribution, each fragment has an excitation energy distribution
having the width of
\begin{equation}
   \delta_{l,h}
   = \frac{\delta_{\rm TXE}}{\sqrt{E_l^2 + E_h^2}} E_{l,h} \ ,
\end{equation}
hence
\begin{equation}
   G(E_x) = \frac{1}{\sqrt{2\pi} \delta_{l,h}}
            \exp\left\{-\frac{(E_x-E_{l,h})^2}{2\delta_{l,h}^2}\right\} \ .
\end{equation}

By creating an initial population $P_0(E_x,J,\Pi) = R(J,\Pi) G(E_x)$
in a compound nucleus, the statistical Hauser-Feshbach calculation is
performed from the highest excitation energy, with a variant version
of the Hauser-Feshbach code CoH$_3$ \cite{Kawano2010}. We include the
neutron and $\gamma$-ray channels only, since the charged particle
emission is strongly suppressed in the neutron-rich nuclei. The
neutron transmission coefficient is calculated by solving the
Schr\"odinger equation for the spherical optical potential, where the
global optical potential parameters of Koning and Delaroche
\cite{Koning2003} are adopted.  The level densities in the continuum
of fission fragments are calculated with the composite level density
formulae of Gilbert and Cameron \cite{Gilbert1965} with an updated
parameterization \cite{Kawano2006}. At low excitation energies, the
discrete level data are taken from the evaluated nuclear structure
database RIPL-3 \cite{RIPL3} (updated in 2012). The $\gamma$-ray
transmission coefficient is calculated from the $\gamma$-ray strength
functions. We adopt the generalized Lorentzian form of Kopecky and Uhl
\cite{Kopecky1990} for the E1 transition with the giant dipole
resonance parameter systematics of Herman {\it et al.}
\cite{INDC0603}.  The higher multipolarities such as the M1 spin-flip
mode and E2 take the standard Lorentzian form with the parameter
systematics in RIPL-3 \cite{RIPL3}.  In addition to the standard
prescription of the $\gamma$-ray strength functions, we also consider
the M1 scissors mode \cite{Mumpower2017}.

A probability of the number of emitted neutrons $P(\nu)$ can be
determined by summing the ground and metastable state production
probabilities of residuals, $P_{Z,A}(\nu)$, where $Z,A$ are for the
fission fragment, $n$ is the number of emitted neutrons. Obviously
$\sum_{\nu} P_{Z,A}(\nu) = 1$ because $P_0(E_x,J,\Pi)$ is normalized.  For
the $k$-th fragment pair of $(Z_l,A_l)$-$(Z_h,A_h)$, $P_{Z_l,A_l}(\nu)$
and $P_{Z_h,A_h}(\nu)$ are calculated separately, and the neutron
multiplicity distribution $P(\nu)$ is calculated by convoluting
them;
\begin{equation}
  P(\nu) = c \sum_{k=1}^N Y_k \sum_{i=0}^{\nu} P_{Z_l,A_l}(i) P_{Z_h,A_h}(\nu-i) \ , 
  \label{eq:Pnu} 
\end{equation}
where $c$ is a normalization constant to satisfy $\sum_{\nu} P_(\nu) =
1$.  Equation (\ref{eq:Pnu}) provides an alternative method to
calculate $\overline{\nu}$, which reads
\begin{equation}
  \overline{\nu} = \sum_{\nu} \nu P(\nu) \ .
\end{equation}

\section{Model Parameter Determination}
\label{sec:Parameter}
\subsection{Determination of the Gaussian parameters}

We determine the five-Gaussian parameters $\Delta_i$, $\sigma_i$, and
$F_i$ in Eq.~(\ref{eq:fivegaussian}) by fitting $Y(A)$ to the
experimental data of Baba {\it et al.} \cite{Baba1997}, Hambsch \cite{Hambsch},
Pleasonton {\it et al.} \cite{Pleasonton1972}, ,
Simon {\it et al.} \cite{Simon1990}, Straede {\it et al.}
\cite{Straede1987}, and Zeynalov {\it et al.} \cite{Zeynalov2006} for
the thermal neutron induced fission on $^{235}$U. These experimental
data are reported as the primary fission fragment yield.

\begin{figure}
  \begin{center}
    \resizebox{\columnwidth}{!}{\includegraphics{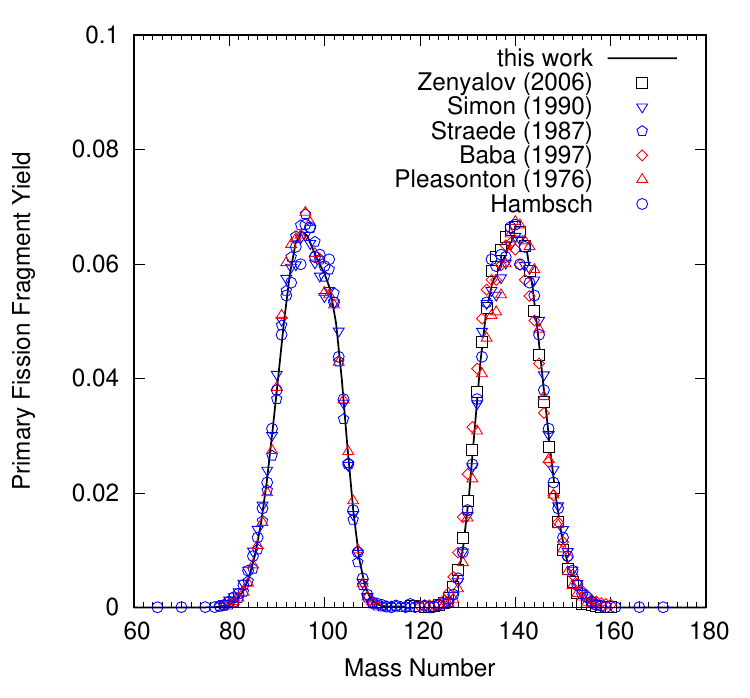}}
  \end{center}
  \caption{The experimental primary fission fragment yield distribution
    $Y(A)$ and the resulting function of Eq.~(\ref{eq:fivegaussian}).}
  \label{fig:yagaussfit}
\end{figure}

Since the mass distribution is symmetric with respect to $A_m$,
relations like $\Delta_1=-\Delta_5$, $\sigma_2=\sigma_4$, hold.  The
normalization condition reads $2(F_1 + F_2) + F_3 = 2$. The obtained
Gaussian parametes are $F_1 = 0.7846$, $\Delta_1 = 23.00$, $\sigma_1 =
4.828$, $F_2 = 0.2032$, $\Delta_2 = 15.63$, $\sigma_2 = 2.728$, $F_3 =
0.002916$, and $\sigma_3 = 8.6$.  Figure~\ref{fig:yagaussfit} shows
the comparison of the five-Gaussian represented $Y(A)$ with the
experimental data.

To predict the independent fission product yields for the
$^{235}$U(n,f) reaction in the fast energy range, we need to estimate
energy variations of the model parameters such as TKE and $Y(A)$ as a
function of neutron incident energy. Because the experimental data of
$Y(A)$ above the thermal energy are very limited, we estimate the
energy dependence of the five-Gaussian parameters solely from the data
of D'yachenko {\it et al.}  \cite{Dyachenko1969}, anchoring the
thermal point determined separately as aforementioned. The fitting is
performed to the data up to the second chance fission threshold, so
that there is no guarantee of extrapolation to the outside region. The
fitted Gaussian parameters, to which we assumed a linear dependence on
the neutron incident energy $E_n$ in MeV, are
\begin{eqnarray}
   \Delta_1 &=& -\Delta_5 = 23.00 - 0.2592 E_n \ , \label{eq:Delta1E} \\
   \Delta_2 &=& -\Delta_4 = 15.63 - 0.1996 E_n \ , \label{eq:Delta2E} \\
   \sigma_1 &=& \sigma_5 = 4.828 + 0.1667 E_n \ ,  \label{eq:sigma1E}
\end{eqnarray}
while $\sigma_2 = \sigma_4$ and $\sigma_3$ are energy independent. The
fractions of each Gaussian are given by
\begin{eqnarray}
   F_2 &=& F_4 = 0.2032 - 0.01565 E_n \ ,   \label{eq:Y2}\\
   F_3 &=& 0.003 + 0.004 E_n \ , \label{eq:Y3}\\
   F_1 &=& F_5 = 1 - F_2 - F_3/2 \ . \label{eq:Y1}
   \label{eq:YE}
\end{eqnarray}

\subsection{Determination of the TKE parameters}

\begin{figure}
  \begin{center}
    \resizebox{\columnwidth}{!}{\includegraphics{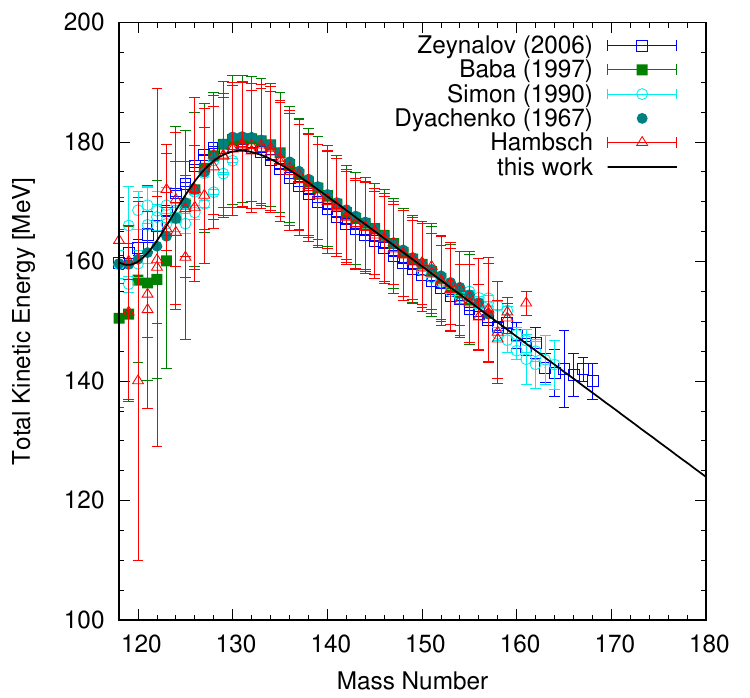}}
  \end{center}
  \caption{The total kinetic energy, TKE, as a function of the heavy fragment
    mass number. The solid line is Eq.~(\ref{eq:TKE}) with the parameters
    given in the text.}
  \label{fig:TKE}
\end{figure}

The parameters of ${\rm TKE}(A)$ in Eq.~(\ref{eq:TKE}) are obtained by
fitting this function to the experimental data of Baba {\it et al.}
\cite{Baba1997}, Hambsch \cite{Hambsch}, Simon {\it et al.}
\cite{Simon1990}, Zeynalov {\it et al.} \cite{Zeynalov2006}, and
D'yachenko {\it et al.} \cite{Dyachenko1969} for the thermal neutron
induced fission on $^{235}$U.  The values of fitting parameters are
$p_1 = 335.3$~MeV, $p_2= 1.174$~MeV, $p_3=0.1876$, and
$p_4=69.08$. $\epsilon_{\rm TKE}$ is slightly adjusted so that
$\overline{\rm TKE}$ is equal to the recommended value of 170.5~MeV
\cite{Wagemans1991}.  The quality of data fitting is demonstrated in
Fig.~\ref{fig:TKE}.

The energy dependece of $\overline{\rm TKE}$ is estimated directly
from experimental data. Madland \cite{Madland2006} estimated a linear
relationship between the neutron incident energy and $\overline{\rm TKE}$. We introduce
an equation similar to Eq~(\ref{eq:TKE}) in order to take
into account a small non-linear tendency seen in the
experimental data in the fast energy region \cite{Lestone2014,
  Duke2014}, which reads
\begin{equation}
  \overline{\rm TKE}(E_n) = (q_1 - q_2 E_n)
               \left\{
                  1 - q_3 \exp\left(-\frac{E_n}{q_4}\right)
               \right\} \ .
  \label{eq:TKE_E}
\end{equation}

\begin{figure}
  \begin{center}
    \resizebox{\columnwidth}{!}{\includegraphics{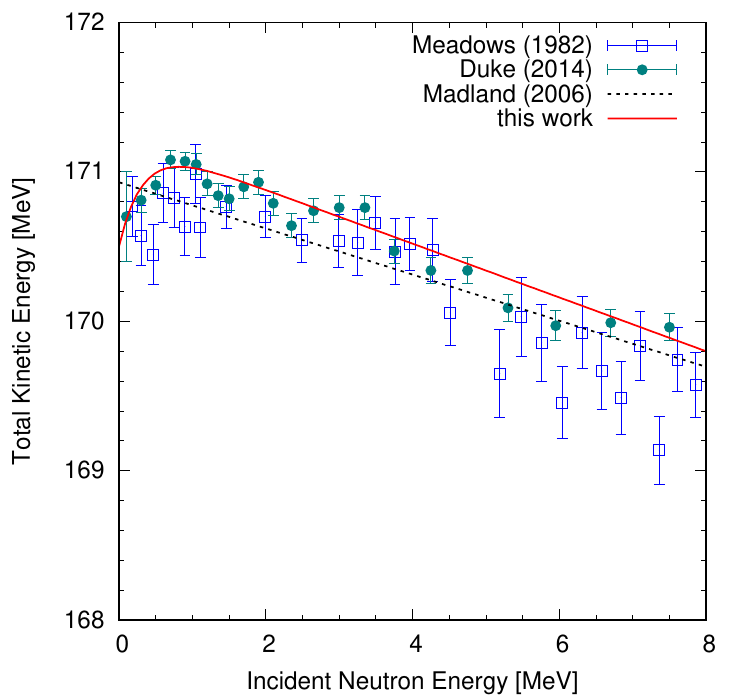}}
  \end{center}
  \caption{The total kinetic energy, TKE, as a function of incident
    neutron energy.  The solid line is Eq.~(\ref{eq:TKE_E}), and the
    dashed line is obtained by Madland \cite{Madland2006}. Meadows'
    \cite{ANLNDM64} data are adjusted to $\overline{\rm TKE}(E_{\rm
      th}) = 170.5$~MeV.}
  \label{fig:TKE_E}
\end{figure}

By fitting Eq.~(\ref{eq:TKE_E}) to the data of Duke \cite{Duke2014},
we obtained $q_1 = 171.2$~MeV, $q_2 = 0.1800$, $q_3 = 0.0043$, and
$q_4 = 0.3230$~MeV.  Figure~\ref{fig:TKE_E} shows the comparison of
this function with the experimental data of Meadows and Budtz-J{\o}rgensen
\cite{ANLNDM64}, and Duke \cite{Duke2014}. At higher energies the
mass-dependence of TKE is assumed to be the same as
Eq.~(\ref{eq:TKE}).

\subsection{Determination of the model parameters in the Hauser-Feshbach calculation}

The scaling factor $f$ in Eq.~(\ref{eq:rjp}) and the anisothermal
parameter $R_T$ in Eq.~(\ref{eq:rt}) are the key ingredients that
control the statistical decay of the fission fragments in the HF$^3$D
model. In our previous study \cite{Kawano2013}, we adopted $R_T=1.3$
and $f=3.0$, which were roughly estimated. We need to revisit these
parameters, since some model inputs such as the Gaussian parameters
were revised.  These parameters should be determined such that the
calculated quantities from the primary fission fragment generated with
the five Gaussians are consistent with many fission observables, which
provides higher confidence in the calculated $Y_I(Z, A)$.

We discretize the anisothermal parameter $R_T$ and the scaling factor
$f$ within some parameter ranges that are physically sound, and study
their impact on some fission observables, $\overline{\nu}$,
$\overline{\nu}(A)$, $P(\nu)$, and $Y_I(A)$. Unfortunately the
experimental data are too uncertain and prevent us to perform a
least-squares fit. However, we were still able to estimate reasonable
values for these parameters.

For the neutron emission multiplicity distribution $P(\nu)$, there
are several experimental data \cite{Boldeman1967, Franklyn1978,
  Diven1956, Holden1988} available to compare with our calculation.
We confirmed that $R_T$ modestly exerts influence upon $P(\nu)$,
while it has a visible sensitivity to the mass-dependent neutron
multiplicities $\overline{\nu}(A)$. We also noticed that the evaluated
$\overline{\nu}$ can be reproduced by adjusting the scaling factor
$f$. When we increase the scaling factor $f$, the spin distribution
$R(J,\Pi)$ of Eq.~(\ref{eq:rjp}) becomes wider, and the system will
have more higher spin populations. Increase in the scaling factor $f$
brings significant reduction in $\overline{\nu}$. The neutron emission
is somewhat hindered due to a spin mismatch between the compound and
daughter nuclei, and it results in an increase in the $\gamma$-ray
emission.  We estimated $f=3.0$ by comparing the evaluated
$\overline{\nu}$ in JENDL-4 of 2.42 at the thermal energy.

\begin{figure}
  \begin{center}
    \resizebox{\columnwidth}{!}{\includegraphics{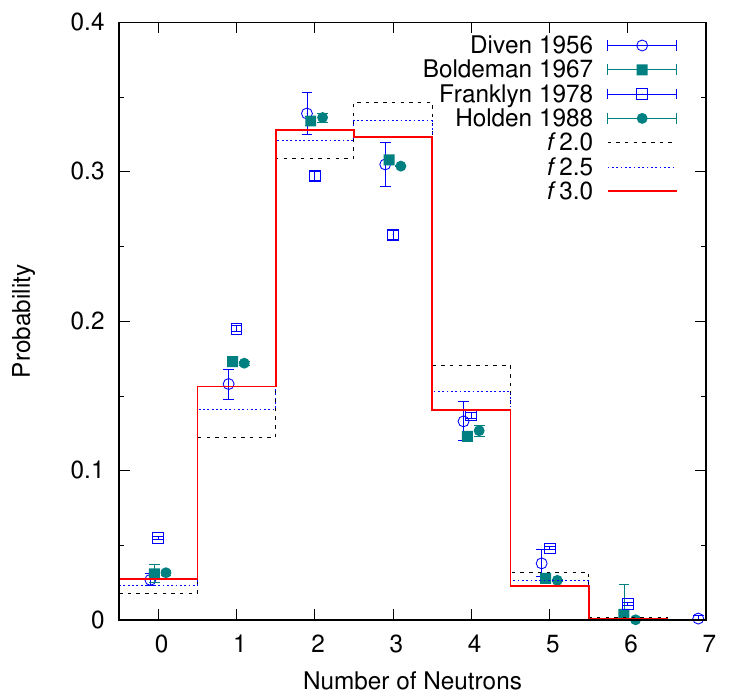}}
  \end{center}
  \caption{The calculated $P(\nu)$ with the anisothermal parameter
    $R_T = 1.2$ and the scaling factor $f = 2.0$, 2.5 and 3.0.}
  \label{fig:Pnu}
\end{figure}

Figure~\ref{fig:Pnu} shows an example of calculated $P(\nu)$ by
changing the scaling factor $f$ while $R_T$ is fixed to 1.2, comparing
with the experimental data \cite{Boldeman1967, Franklyn1978,
  Diven1956, Holden1988}. Based on this exercise together with the
calculated value of $\overline{\nu}$, we adopted $f$ = 3.0, which
gives a relatively good agreement with the experimental data.

\begin{figure}
  \begin{center}
    \resizebox{\columnwidth}{!}{\includegraphics{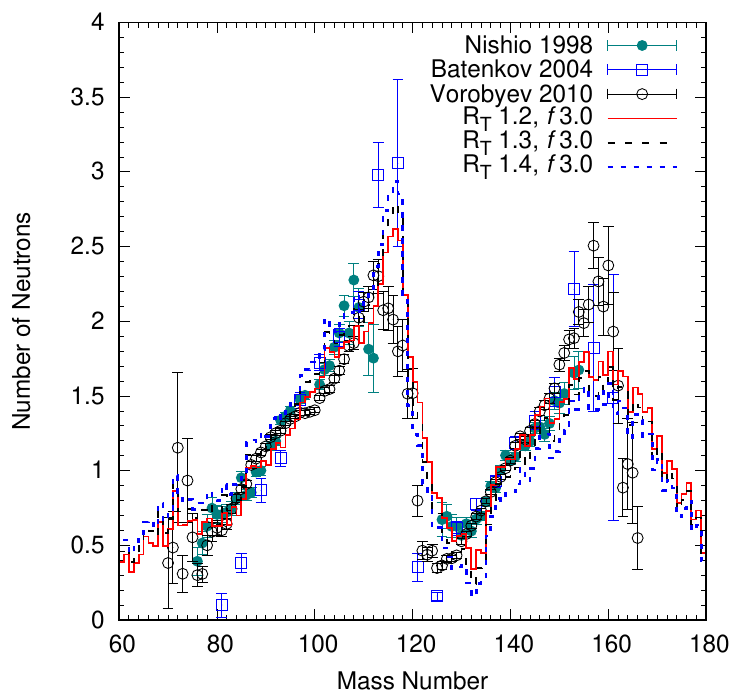}}
  \end{center}
  \caption{The calculated $\overline{\nu}(A)$ with the scaling factors $f$ =
    3.0 and the anisothermal parameter $R_T$ = 1.2, 1.3 and 1.4.}
  \label{fig:nuA}
\end{figure}

An impact of $R_T$ on the mass-dependent average neutron multiplicity
$\overline{\nu}(A)$ is shown in Fig.~\ref{fig:nuA}, where the cases of
$R_T = 1.2$, 1.3 and 1.4 are shown. We have already revealed that the
$Y(Z, A, TKE)$ is the most important input to reproduce the sawtooth
structure of $\overline{\nu}(A)$ \cite{Talou2011}. Small adjustment
can be possible by varying $R_T$. The scaling factor $f$ is fixed at
3.0 where the best fit of $\overline{\nu}$ can be reproduced. Despite
our observation that $R_T$ does not change the shape of $P(\nu)$,
$\overline{\nu}(A)$ is notably affected by $R_T$. Generally speaking,
increase in $R_T$ gives a modest change in $\overline{\nu}$ and
$P(\nu)$ at the thermal energy.

According to Eq. (8), more excitation energy is given to the light
fragments when $R_T>1$, and the number of neutrons emitted from the
light fragment increases. We noticed that a value of $R_T = 1.2$
reproduces the experimental data most reasonably.  The measurement of
$\overline{\nu}(A)$ could be very difficult in the whole fission
product mass range, since the yield can be extremely small in some
mass regions, and it obliges the measurement severe statistics.
Because the deterministic method mitigates such a difficult condition,
we emphasize that our calculated result is not just an estimation but
prediction supported by consistent descriptions of several observables
simultaneously.

\begin{figure}
  \begin{center}
    \resizebox{\columnwidth}{!}{\includegraphics{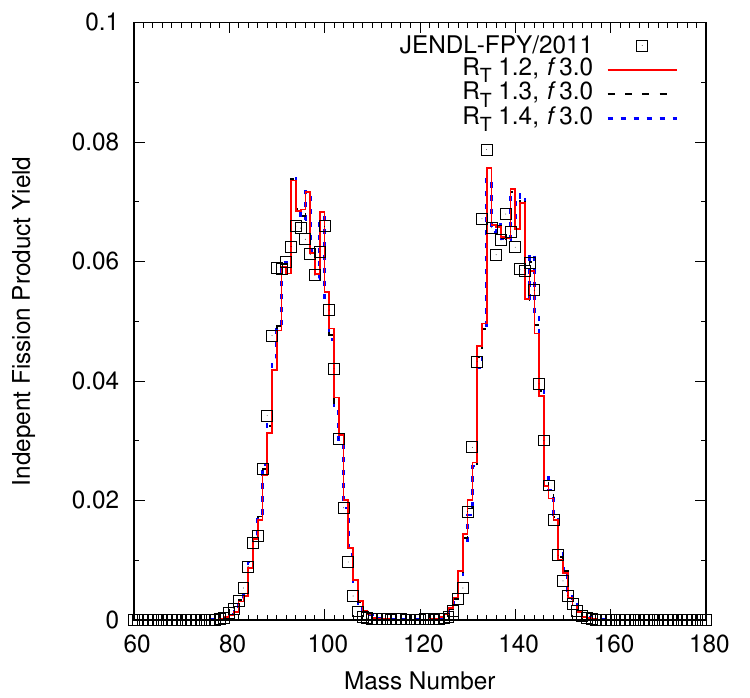}}
  \end{center}
  \caption{The calculated post-neutron distribution of fission products
    $Y_I(A)$ with the scaling factor $f$ = 3.0 and the anisothermal
    parameter $R_T$ = 1.0, 1.2 and 1.4.}
  \label{fig:ya}
\end{figure}

These parameters also change the calculated mass chain yield of the
independent FPY, $Y_I(A)$.  We carried out comparisons of $Y_I(A)$
with the ones in the evaluated libraries by varying the $R_T$ and $f$
parameters.  A representative result for $Y_I(A)$ with the parameters
$R_T = 1.0$, 1.2 and 1.4 (the scaling factors $f = 3.0$) is shown in
Fig.~\ref{fig:ya}. The mass distribution somehow extends outward with
increasing $R_T$.  Similar to the case of $\overline{\nu}(A)$, $R_T =
1.2$ provides a reasonable agreement with the evaluated data.
Considering the arguments in this section, we conclude $R_T = 1.2$ and
$f = 3.0$ to be the best set of the constants in this study.

Figure~\ref{fig:ya} also indicates that the Hauser-Feshbach
calculation for the neutron evaporation process successfully
reproduces the post-neutron mass distribution $Y_I(A)$. The
pre-neutron distribution $Y(A)$ represented by the Gaussians is
symmetric with respect to $A_m$, and the symmetry is broken after the
prompt neutron emission. The evaluated mass distribution of
independent FPY, such as those in ENDF/B-VI and JENDL/FPY-2011,
exhibits some noticeable yield peaks at $A = 93$, 96, and 99 in the
light fragment and $A = 133$, 134 and 138 in the heavy fragment. The
origin of some of these peaks can be understood from
$\overline{\nu}(A)$. For instance the peaks at $A = 133$ and 134 can
be related to the doubly magic nucleus at $A = 132$ appeared as a
sudden drop in $\overline{\nu}(A)$.  Similarly other peaks are formed
due to the excitation energy sorting mechanism and the difference in
the neutron separation energies.  Our calculation reproduces the peaks
at $A = 93$, 96 and 134.

\section{Results and Discussion}
\label{sec:Results}
\subsection{Calculated independent fission product yield}

\begin{figure}
  \begin{center}
    \resizebox{.7\columnwidth}{!}{\includegraphics{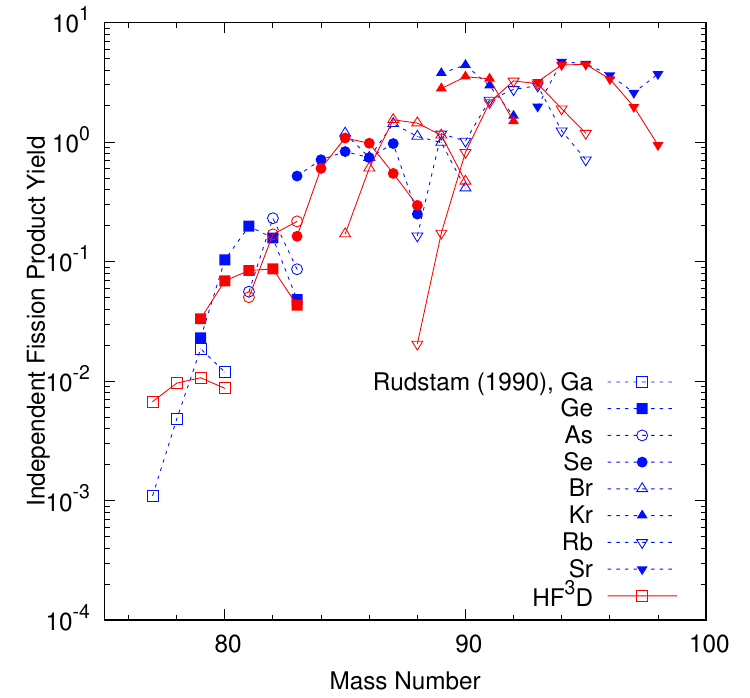}}\\
    \resizebox{.7\columnwidth}{!}{\includegraphics{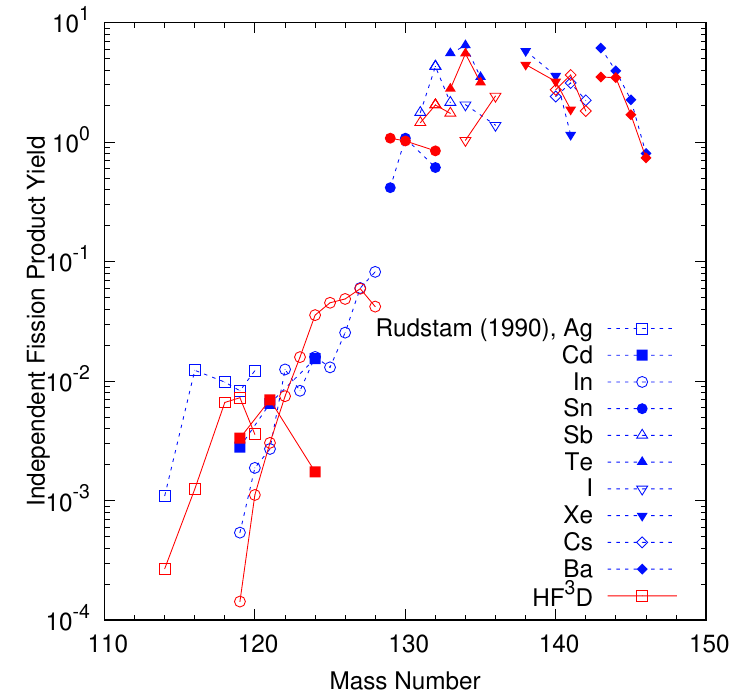}}
  \end{center}
  \caption{Comparison of the calculated independent yield $Y_I(Z,A)$ with
    the experimental data reported by Rudstam {\it et al.} \cite{Rudstam1990}}
  \label{fig:yindep}
\end{figure}

We compare our calculated independent yield with a limited number of
experimental data. This is because the present HF$^3$D model is more
or less the proof of concept to demonstrate the feasibility of fully
deterministic calculations for FPY. The calculated $Y_I(Z,A)$ for
several fission products are compared with the experimental data of
Rudstam {\it et al.} \cite{Rudstam1990} in
Fig.~\ref{fig:yindep}. Although HF$^3$D is not able to fit the data
precisely, general tendency is well reproduced. The C/E-value
(Calculated/Experimental) varies from 0.10 ($^{116}$Ag) to 6.12
($^{77}$Ga) and the average is $1.03\pm0.81$. This is partly because
Wahl's $Z_p$ model works reasonably well. Note that the experimental
$Y_I(Z,A)$ data are not the direct measurements but derived from the
cumulative data \cite{Rudstam1987}.

\subsection{Energy dependent result}

\begin{figure}
  \begin{center}
    \resizebox{\columnwidth}{!}{\includegraphics{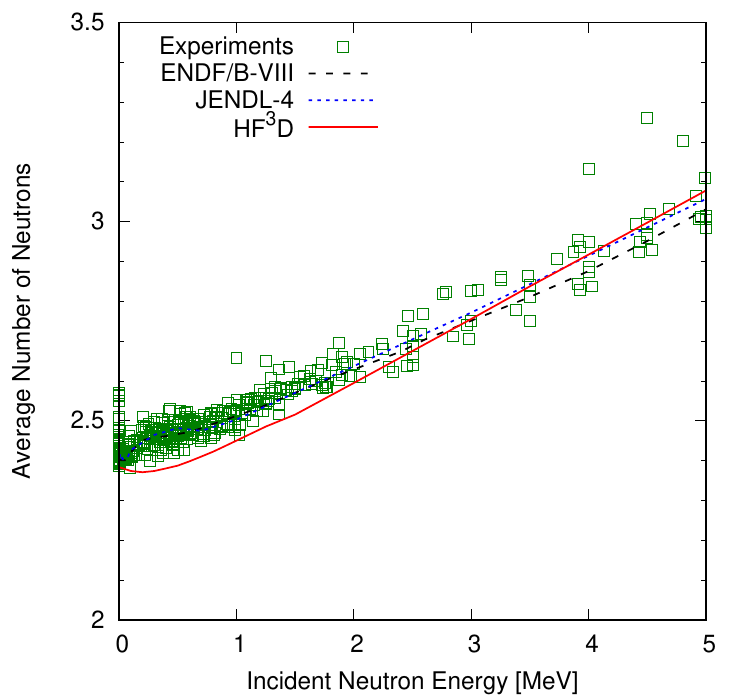}}
  \end{center}
  \caption{The calculated average number of prompt neutrons
    $\overline{\nu}$ comparing with the evaluated $\overline{\nu}$ in
    ENDF/B-VIII (dashed line) and JENDL-4 (dotted line), and
    experimental data taken from the EXFOR databaes.}
  \label{fig:nubar_e}
\end{figure}

To confirm the estimated incident energy dependence in $Y(A)$ and
$\overline{\rm TKE}$ in the HF$^3$D model, we compare $\overline{\nu}$
with the evaluated values at energies above thermal.  $\overline{\nu}$
tells us information on an energy balance amongst the total fission
energy, TKE, the kinetic energy of evaporated neutrons, and the
emitted $\gamma$-rays.  We assume that the anisothermal parameter
$R_T$ = 1.2 and the scaling factor $f$ = 3.0 determined at the thermal
energy do not change, at least up to the second chance fission
threshold.  The calculated $\overline{\nu}$ is shown in
Fig.~\ref{fig:nubar_e}, which compares with the evaluated
$\overline{\nu}$ in ENDF/B-VIII.0 and JENDL-4 \cite{JENDL4} together
with experimental data taken from EXFOR database.  Our calculated
values are fairly consistent with these evaluations, although slightly
deviate below 2~MeV.  The deviations from the evaluated values at
1~MeV are 2.1\% (JENDL-4) and 2.5\% (ENDF/B-VIII.0), respectively.

The mass-dependent average neutron multiplicity $\overline{\nu}(A)$ in
Fig.~\ref{fig:nuA} increases as the incident neutron energy.  The
changes in $\overline{\nu}(A)$ at each $A$ are roughly uniform hence
we do not show a figure, except in the mass regions near $A_l=100$ and
$A_h=150$ where the rise seems to be slightly larger. It is reported
that an extra energy is always transferred to the heavy fragments when
the incident energy increases \cite{Muller1984}.  The current model,
however, predicts increases in $\overline{\nu}(A)$ from both the light
and heavy fragments, because we assumed $R_T$ is constant.  If $R_T$
decreases as the incident neutron energy increases, the change in
$\overline{\nu}(A)$ would be more prominent in the heavy fragment
side. At this moment we have no clear reason and strong evidence to
explain the reduction in $R_T$.
It is also noted that the the mass region around $A = 118$ generated
mostly by the symmetric fission mode that is expresed by $R_T =
1.0$. Due to very low yields around $A = 118$, we ignored this effect.

\begin{figure}
  \begin{center}
    \resizebox{\columnwidth}{!}{\includegraphics{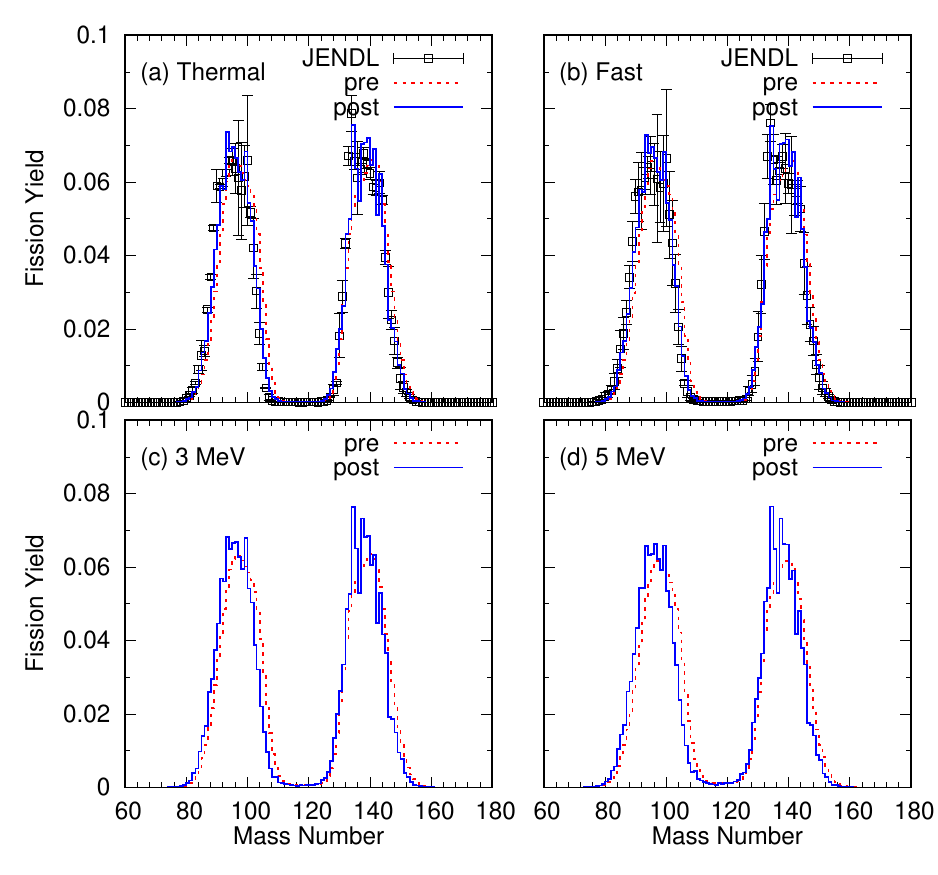}}
  \end{center}
  \caption{The calculated $Y_I(A)$ at (a) thermal, (b) fast energy,
    (c) 3, and (d) 5~MeV.  The evaluated $Y_I(A)$ in JENDL/FPY-2011 at
    the thermal and fast energies are also depicted in the top
    panels.}
  \label{fig:ya_e}
\end{figure}

The distribution of fission fragment yield evolves from the primary
$Y(A)$ (pre-neutron) to the independent yield $Y_I(A)$ (post-neutron)
distributions, which are connected by the prompt neutron emission
calculated with the Hauser-Feshbach theory.  Figure~\ref{fig:ya_e}
shows the comparison of the mass distributions for the pre- and
post-neutron emission at four energies, together with the independent
mass yields in JENDL-FPY/2011 at the thermal and fast energies.
Obviously the difference between the thermal and fast energies is
modest, since their energies are close each other. As the energy
increases, the post-neutron distribution shifts toward the low-$A$
side.  This shift is larger in the mass regions of 100 and 150.

\subsection{Evaluation of Isomeric Ratio}

Radiochemical determination of independent yield ratios of isomers of
known spins for $^{235}$U(n,f) has been conducted for some of major
fission products, {\it e.g.}  $^{128,130,132}$Sb, $^{131,133}$Te,
$^{132,134,136}$I, $^{135}$Xe and $^{138}$Cs, and summarized by Naik {\it
  et al.} \cite{Naik1995}.  The reported partial yields of ground and
isomeric states were determined by the cumulative yield after
correcting the precursor contributions. The isomeric ratio is defined as
\begin{equation}
  {\rm IR} = \frac{Y_m}{Y_m+ Y_g}  \ ,
  \label{eq:ir}
\end{equation}
where $Y_g$ and $Y_m$ are the partial yield of ground and isomeric
states in a specific nuclide. Because measurements of the isomeric
ratios for all the fission products often encounter technical
difficulties, theoretical predictions are essential for evaluating
the nuclear data files. A model widely used for evaluating the
isomeric ratio data was first proposed by Madland and England
\cite{Madland1977,LA-6595-MS}. In the Madland-England (ME) model the
average angular momentum of the initial fission fragment is considered
as a model parameter. The spin distribution of the fragments after
prompt neutron emission is
\begin{equation}
  P(J) = P_0(2J+1)
        \exp\left[
              -\frac{(J+1/2)^2}{\langle J^2\rangle}
            \right] \ ,
\label{eq:angdist}
\end{equation}
where $P_0$ is the normalization constant, $J_{rms} \equiv
\sqrt{\langle J^2\rangle}$ characterizes the angular momentum of an
initial fragment.  As given in Eq.~(\ref{eq:angdist}), the fission
fragments are assumed to be formed with a density distribution $P(J)$
of the total angular momentum $J$, which is parameterized by $J_{rms}$.
The parameter $J_{rms}$ is assumed to be constant for all fission
fragment masses in the neutron induced fission, whereas $J_{rms}$ 
varies with incident neutron energy. The model adopted $J_{rms} = 7.5$ for
the thermal neutron induced fission for all the fission fragments.

The model gives IR or the branching ratio ($Y_m/Y_g$) for eight
different cases, whether the fission product mass number is even or
odd, whether the spin difference between the metastable ($J_m$) and
ground state ($J_g$) $\mid J_m-J_g \mid$ is even or odd, and whether
$J_m$ is greater or less than $J_g$. Since predicted isomeric ratio by
this model depends on $J_{m/g}$ and $A$ only, all the nuclides having
the same $J_{m/g}$ and even/odd $A$ will have the same isomeric ratio.
For example, the ME model gives the same isomeric ratio of IR = 0.707
for both $^{133}$Xe and $^{135}$Xe, nevertheless the experimental data
\cite{Ford1984} indicate they differ. In addition, due to the inherent
simplification in the model, the accuracy of the estimated branching
ratio could be limited, particularly when $J_m$ is very high compared
to $J_g$, or there are other possible metastable states.  When a
nuclide has more than one metastable state, the definition of IR must
be corrected as $Y_{m}^{(i)}/(Y_g + \sum_k Y_{m}^{(k)})$.

\begin{figure}
  \begin{center}
    \resizebox{\columnwidth}{!}{\rotatebox{90}{\includegraphics{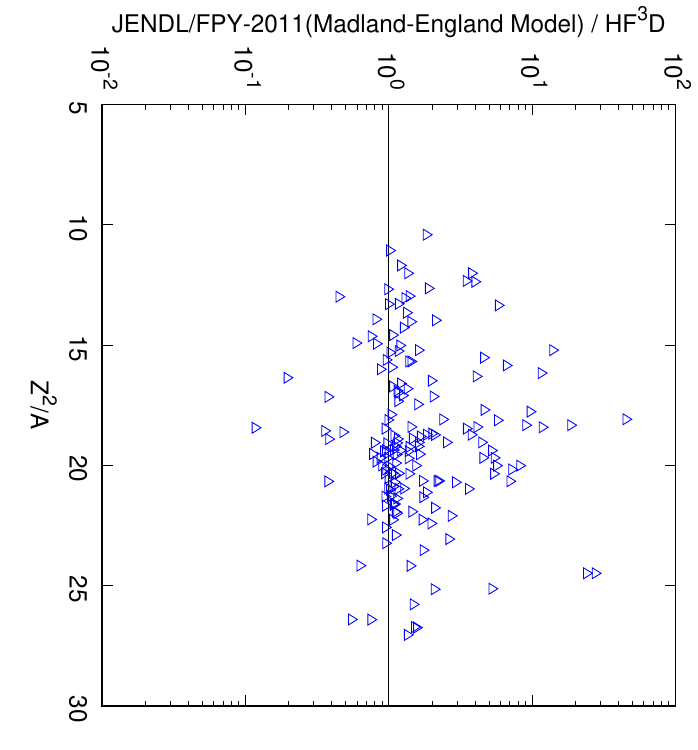}}}
  \end{center}
  \caption{The ratio of the Madland-England model to the Hauser-Feshbach
    calculation by HF$^3$D for all possible fission products containing the metastable
    state.}
  \label{fig:coe}
\end{figure}

We performed the HF$^3$D model calculation for many primary fission
fragments, and searched for all the levels in RIPL-3 \cite{RIPL3}
whose half-life is longer than 1~ms. This definition could be somewhat
shorter than the half-life of commonly known metastable states.  We
paid special attention to include some high-lying metastable states.
If its excitation energy is higher than the known level up to which
RIPL-3 says there is no missing level, we postulates several levels
between these states according to the CoH$_3$ level density and
spin-parity distributions to fill the gap.  Figure~\ref{fig:coe} shows
the ratio of calculated IRs with the ME model (JENDL/FPY-2011) to
those with HF$^3$D, plotted against $Z^2/A$. The mean value of the
ratio 2.75 $\pm$ 4.90 indicates a clear disagreement between the ME
and HF$^3$D models, and the ME model tends to overestimate IR.

\begin{figure*}
  \begin{center}
    \resizebox{1.0\textwidth}{!}{\rotatebox{90}{\includegraphics{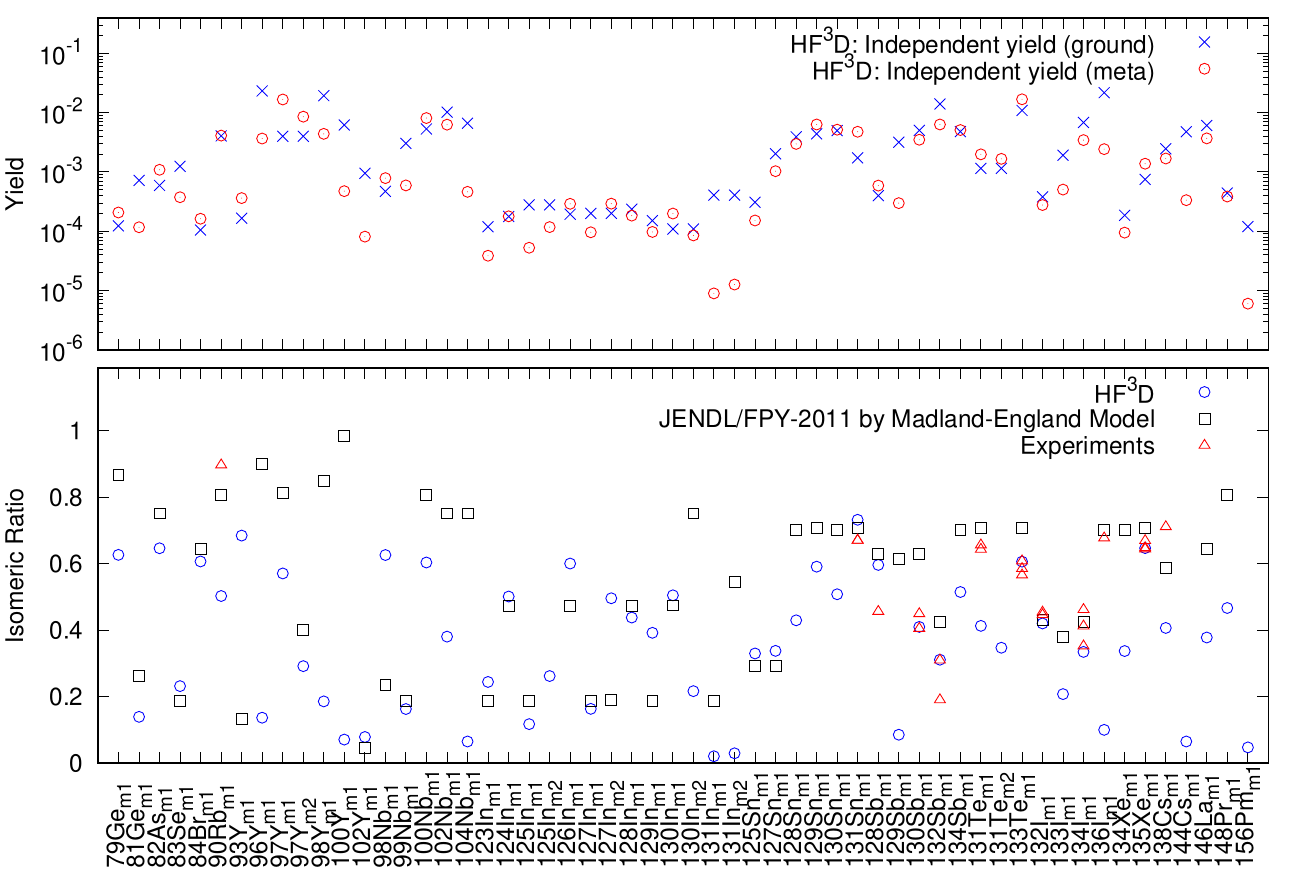}}}
  \end{center}
  \caption{Calculated fractional independent yields (upper panel) and
    IRs (bottom panel) for the nuclides that have the ground state
    yield of greater than $1 \times 10^{-4}$. The IRs are shown together with
    JENDL/FPY-2011 data compiled by the Madland-England model, and
    some experimental data. The first and the second isomeric states
    are denoted by the subscript.}
  \label{fig:ir}
\end{figure*}

The disagreement is also evident in Figure~\ref{fig:ir}, where IRs for
each individual independent fission product, whose yield is more than
$10^{-4}$, are shown. The upper panel shows the calculated fractional
independent yields for the ground and metastable states, and the lower
panel shows their IRs together with some experimental IRs
\cite{Reeder1985, Naik1995}. Although the experimental IR data are scarce, we can
compare some of the results for $^{90}$Rb, $^{131}$Sn, $^{128, 130,
  132}$Sb, $^{131,133}$Te, $^{132,134,136}$I, $^{133,135}$Xe and
$^{138}$Cs.

One of the interesting features in our model is that we find two
isomeric states in $^{131}$Te. The experimental IR was reported in
1995 and the second metastable state with $J^{\pi} = (23/2)^+$ was
reported in 1998 \cite{Fogelberg1998}, and both ENDF/B-VII and
JENDL/FPY-2011 only include the first metastable state so far. This
means the reported IR is understood to be
$(Y_{m1}+Y_{m2})/(Y_g+Y_{m1}+Y_{m2})$, while we calculated this as
$Y_{m1}/(Y_g+Y_{m1}+Y_{m2})$ and $Y_{m2}/(Y_g+Y_{m1}+Y_{m2})$
independently. This is why our calculation looks lower than the
experimental data. In fact our calculated
$(Y_{m1}+Y_{m2})/(Y_g+Y_{m1}+Y_{m2})$ value agrees fairly well with
the experimental data.

Besides $^{134}$I and $^{136}$I, the ME model better agrees
with the experimental data than our model.  We examined if some
even/odd effects in the spin and mass number reveal systematic
features in the ME prediction. However no prominent rule was found.

In the case of $^{128,130}$Sb, where $Jg$ is higher than $J_m$, our
model predicts IR pretty reasonably. That said, our calculation misses
the experimental data of $^{136}$I, where the difference in $J_g$ and
$J_m$ is relatively large.  At this moment we have no simple
explanation of why our prediction fails in some cases, although there
are not so many. Missing high spin levels in the RIPL data file can
cause such discrepancies and thus it requres the better nuclear structure
data.

\begin{figure}
  \begin{center}
\resizebox{\linewidth}{!}{\includegraphics{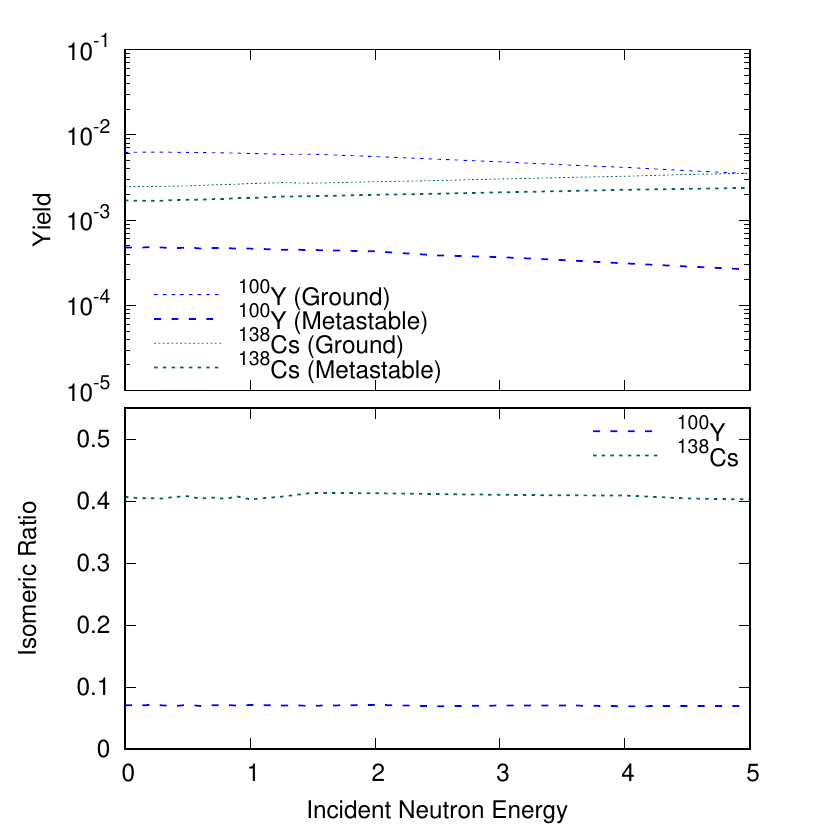}}
  \end{center}
  \caption{Energy dependence of ground and metastable state yield (top frame) and
    isomeric ratio (bottom frame) of $^{100}$Y and $^{138}$Cs.}
  \label{fig:ir_e}
\end{figure}

The energy dependence of IR for is confirmed. It is revealed that
IR is insensitive to the incident neutron energy in general. In
Figure~\ref{fig:ir_e}, $^{100}$Y and $^{138}$Cs are chosen as
examples. While the fractional independent yield of ground and
metastable states change slightly as the incident neutron energy
incereases, the IRs remain constant.

\section{Conclusion}
\label{sec:Conclusion}
We developed a new method to calculate the independent fission product
yield $Y_I(Z, A)$ and the isomeric ratio IR by applying the
statistical Hauser-Feshbach theory to the decay process ofthe primary
fission fragment pairs for $^{235}$U(n$_{\rm th}$,f), where about
1,000 nuclides are involved. Instead of employing the Monte Carlo
sampling technique that was commonly used in the past, our model,
called HF$^3$D, is based on the fully deterministic technique.  The
input data, {\it e.g.} the mass distribution of the primary fission
fragment and the fragment kinetic energy were determined by fitting
analytical functions to the available experimental data, the charge
distribution was generated by the $Z_p$ model, and the incident neutron
energy dependence was also determined according to the experimental
data.  Besides these inputs, we adjusted the scale factor $f$ in the
spin distribution and the anisothermal parameter $R_T$, which define
the initial configuration of the fission fragments, to reproduce some
observables in fission, such as the average prompt neutron
multiplicity $\overline{\nu}$, its mass dependence
$\overline{\nu}(A)$, and the neutron emission multiplicity
distribution $P(\nu)$.

Beginning with the initial configuration characterized by the primary
fission fragment pair, the total kinetic energy, the excitation
energy, and the initial spin and parity distributions, we calculated
the neutron evaporation both from the light and heavy fragments of
$^{235}$U(n,f). We demonstrated that the anisothermal parameter $R_T$
is sensitive to $\overline{\nu}(A)$, while the initial distribution
controlled by the scale factor $f$ impacts on $P(\nu)$. Once these
parameters were tuned, we showed that the calculated post-neutron mass
distribution $Y_I(A)$ agreed well with that in JENDL/FPY-2011.

We demonstrated that the symmetric mass distribution of the primary
fission fragment $Y(A)$ with respect to $A_m= 236/2 = 118$ is broken
due to the prompt neutron emission, and the resulting $Y_I(A)$
evidently shows the mass peaks at $A = 93$, 96 and 134 seen also in the
evaluated FPY data.

In addition to $Y_I(Z,A)$, the model also provides the IR for the
nuclides having any short-lived or long-lived isomeric states.  We
observed that IRs predicted by the Madland-England
model \cite{Madland1977,LA-6595-MS} tend to be larger than our
calculation.

Optimizing our model input parameters on the thermal neutron data, we extrapolated the
model calculations up to 5~MeV, where only the first chance fission
takes place.  Generally speaking IRs are insensitive to the incident
neutron energy.

As a final remark, by connecting the HF$^3$D method with the decay
data library, we will be able to predict the cumulative fission
product yield as well as the mass chain yield, for which abundant
experimental data are accessible. This would be a powerful tool to
evaluate the independent and cumulative fission product yield data
simultaneously and consistently in the energy-dependent manner, and
obviously be our next stride.

\section{Acknowledgements}
We are grateful to T. Yoshida of Tokyo Institute of Technology (Tokyo
Tech.) and K. Nishio of Japan Atomic Energy Agency (JAEA) for valuable
discussions. This work was partially supported by the grant
``Development of prompt-neutron measurement in fission by surrogate
reaction method and evaluation of neutron-energy spectra,'' entrusted
to JAEA by Ministry of Education, Culture, Sports, Science and
Technology. This work was partly carried out under the auspices of the
National Nuclear Security Administration of the U.S. Department of
Energy at Los Alamos National Laboratory under Contract
No. DE-AC52-06NA25396.  One of the authors (TK) would like to thank
the WRHI (World Research Hub Initiative) program at Tokyo Tech. for
supporting this work.


\begin{thebibliography}{10}

\bibitem{LA-UR-94-3106}
England TR, Rider BF.
\newblock Evaluation and compilation of fission product yields.
\newblock Los Alamos National Laboratory; 1994. ENDF-349, LA-UR-94-3106.

\bibitem{Yoshida1981}
Yoshida T, Nakasima R.
\newblock Decay Heat Calculations Based on Theoretical Estimation of Average
  Beta- and Gamma-Energies Released from Short-Lived Fission Products.
\newblock Journal of Nuclear Science and Technology. 1981;18(6):393 -- 407.
\newblock Available from: \url{https://doi.org/10.1080/18811248.1981.9733273}.

\bibitem{Yoshida2008}
Yoshida T, Wakasugi Y, Hagura N.
\newblock Pandemonium Problem in Fission-Product Decay Heat Calculations
  Revisited.
\newblock Journal of Nuclear Science and Technology. 2008;45(8):713 -- 717.
\newblock Available from:
  \url{http://www.tandfonline.com/doi/abs/10.1080/18811248.2008.9711471}.

\bibitem{ChibaGo2017}
Chiba G.
\newblock Consistent adjustment of radioactive decay and fission yields data
  with measurement data of decay heat and $\beta$-delayed neutron activities.
\newblock Annals of Nuclear Energy. 2017;101(Supplement C):23 -- 30.
\newblock Available from:
  \url{http://www.sciencedirect.com/science/article/pii/S0306454916305849}.

\bibitem{Yoshida2002}
Yoshida T, Okajima S, Sakurai T, Nakajima K, Yamane T, Katakura Ji, et~al.
\newblock Evaluation of Delayed Neutron Data for JENDL-3.3.
\newblock Journal of Nuclear Science and Technology. 2002;39(sup2):136 -- 139.
\newblock Available from: \url{https://doi.org/10.1080/00223131.2002.10875059}.

\bibitem{ChibaGo2013}
Chiba G, Tsuji M, Narabayashi T.
\newblock Sensitivity and uncertainty analysis for reactor stable period
  induced by positive reactivity using one-point adjoint kinetics equation.
\newblock Journal of Nuclear Science and Technology. 2013;50(12):1150 -- 1160.
\newblock Available from: \url{https://doi.org/10.1080/00223131.2013.838332}.

\bibitem{Hayes2015}
Hayes AC, Friar JL, Garvey GT, Ibeling D, Jungman G, Kawano T, et~al.
\newblock Possible origins and implications of the shoulder in reactor neutrino
  spectra.
\newblock Phys Rev D. 2015 Aug;92:033015.
\newblock Available from:
  \url{http://link.aps.org/doi/10.1103/PhysRevD.92.033015}.

\bibitem{Shibagaki2016}
Shibagaki S, Kajino T, J MG, Chiba S, Nishimura S, Lorusso G.
\newblock Relative Contributions of the Weak, Main, and Fission-recycling
  r-process.
\newblock The Astrophysical Journal. 2016;816(2):79.
\newblock Available from: \url{http://stacks.iop.org/0004-637X/816/i=2/a=79}.

\bibitem{Chadwick2010}
Chadwick MB, Kawano T, Barr DW, Mac~Innes MR, Kahler AC, T G, et~al.
\newblock Fission Product Yields from Fission Spectrum n+$^{239}$Pu for
  ENDF/B-VII.1.
\newblock Nuclear Data Sheets. 2010;111(12):2923 -- 2964.
\newblock Available from:
  \url{http://www.sciencedirect.com/science/article/pii/S009037521000102X}.

\bibitem{Kawano2013b}
Kawano T, Chadwick MB.
\newblock Estimation of $^{239}$Pu independent and cumulative fissio product
  yields from the chain yield data using a Bayesian technique.
\newblock Journal of Nuclear Science and Technology. 2013;50(10):1034 -- 1042.
\newblock Available from: \url{https://doi.org/10.1080/00223131.2013.830580}.

\bibitem{LA13928}
Wahl AC.
\newblock Systematics of Fission-Product Yields.
\newblock Los Alamos National Laboratory; 2002. LA-13928.

\bibitem{JAEA-DATACODE-2011-025}
Katakura J.
\newblock JENDL FP Decay Data File 2011 and Fission Yields Data File 2011.
\newblock Japan Atomic Energy Agency; 2012. JAEA-Data/Code 2011-025.

\bibitem{Katakura2016}
Katakura J, Minato K F~Ohgama.
\newblock Revision of the JENDL FP Fission Yield Data.
\newblock EPJ Web of Conferences. 2016;111:08004--1 -- 5.
\newblock Available from:
  \url{https://www.epj-conferences.org/articles/epjconf/abs/2016/06/epjconf_wonder2016_08004/epjconf_wonder2016_08004.html}.

\bibitem{Madland1977}
Madland DG, England TR.
\newblock The Influence of Isomeric States on Independent Fission Product
  Yields.
\newblock Nuclear Science and Engineering. 1977;64:859 -- 865.
\newblock Available from: \url{https://doi.org/10.13182/NSE77-A14501}.

\bibitem{LA-6595-MS}
Madland DG, England TR.
\newblock Distribution of independent fission-product yields to isomeric
  states.
\newblock Los Alamos National Laboratory; 1994. LA-6595-MS.

\bibitem{Kawano2013}
Kawano T, Talou P, Stetcu I, Chadwick MB.
\newblock Statistical and evaporation models for the neutron emission energy
  spectrum in the center-of-mass system from fission fragments.
\newblock Nuclear Physics A. 2013;913(2):51 -- 70.
\newblock Available from:
  \url{http://www.sciencedirect.com/science/article/pii/S0375947413005952}.

\bibitem{Stetcu2013}
Stetcu I, Talou P, Kawano T, Jandel M.
\newblock Isomer production ratios and the angular momentum distribution of
  fission fragments.
\newblock Phys Rev C. 2013 Oct;88:044603.
\newblock Available from:
  \url{http://link.aps.org/doi/10.1103/PhysRevC.88.044603}.

\bibitem{Becker2013}
Becker B, Talou P, Kawano T, Danon Y, Stetcu I.
\newblock Monte Carlo Hauser-Feshbach predictions of prompt fission
  $\ensuremath{\gamma}$ rays: Application to ${n}_{\mathrm{th}}+{}^{235}$U,
  ${n}_{\mathrm{th}}+{}^{239}$Pu, and ${}^{252}$Cf (sf).
\newblock Phys Rev C. 2013 Jan;87:014617.
\newblock Available from:
  \url{http://link.aps.org/doi/10.1103/PhysRevC.87.014617}.

\bibitem{Litaize2010}
Litaize O, Serot O.
\newblock Investigation of phenomenological models for the Monte Carlo
  simulation of the prompt fission neutron and $\ensuremath{\gamma}$ emission.
\newblock Phys Rev C. 2010 Nov;82:054616.
\newblock Available from:
  \url{https://link.aps.org/doi/10.1103/PhysRevC.82.054616}.

\bibitem{Randrup2009}
Randrup J, Vogt R.
\newblock Calculation of fission observables through event-by-event simulation.
\newblock Phys Rev C. 2009 Aug;80:024601.
\newblock Available from:
  \url{https://link.aps.org/doi/10.1103/PhysRevC.80.024601}.

\bibitem{Randrup2014}
Randrup J, Vogt R.
\newblock Refined treatment of angular momentum in the event-by-event fission
  model freya.
\newblock Phys Rev C. 2014 Apr;89:044601.
\newblock Available from:
  \url{https://link.aps.org/doi/10.1103/PhysRevC.89.044601}.

\bibitem{Schmidt2016}
Schmidt KH, Jurado B, Amouroux C, Schmitt C.
\newblock General Description of Fission Observables: GEF Model Code.
\newblock Nuclear Data Sheets. 2016;131:107 -- 221.
\newblock Special Issue on Nuclear Reaction Data.
\newblock Available from:
  \url{http://www.sciencedirect.com/science/article/pii/S0090375215000745}.

\bibitem{Talou2017}
Talou P, Vogt R, Randrup J, Rising ME, Pozzi SA, Nakae L, et~al.
\newblock Correlated Prompt Fission Data in Transport Simulations.
\newblock European Physical Journal. 2017;.

\bibitem{Aritomo2013}
Aritomo Y, Chiba S.
\newblock Fission process of nuclei at low excitation energies with a Langevin
  approach.
\newblock Phys Rev C. 2013 Oct;88:044614.
\newblock Available from:
  \url{https://link.aps.org/doi/10.1103/PhysRevC.88.044614}.

\bibitem{Aritomo2014}
Aritomo Y, Chiba S, Ivanyuk F.
\newblock Fission dynamics at low excitation energy.
\newblock Phys Rev C. 2014 Nov;90:054609.
\newblock Available from:
  \url{https://link.aps.org/doi/10.1103/PhysRevC.90.054609}.

\bibitem{Usang2016}
Usang MD, Ivanyuk FA, Ishizuka C, Chiba S.
\newblock Effects of microscopic transport coefficients on fission observables
  calculated by the Langevin equation.
\newblock Phys Rev C. 2016;94:044602.
\newblock Available from:
  \url{https://link.aps.org/doi/10.1103/PhysRevC.94.044602}.

\bibitem{Ishizuka2017}
Ishizuka C, Usang MD, Ivanyuk FA, Maruhn JA, Nishio K, Chiba S.
\newblock Four-dimensional Langevin approach to low-energy nuclear fission of
  $^{236}$U.
\newblock Phys Rev C. 2017;96:064616.
\newblock Available from:
  \url{https://link.aps.org/doi/10.1103/PhysRevC.96.064616}.

\bibitem{Usang2017}
Usang MD, Ivanyuk FA, Ishizuka C, Chiba S.
\newblock Analysis of the total kinetic energy of fission fragments with the
  Langevin equation.
\newblock Phys Rev C. 2017;96:064617.
\newblock Available from:
  \url{https://link.aps.org/doi/10.1103/PhysRevC.96.064617}.

\bibitem{Sierk2017}
Sierk AJ.
\newblock Langevin model of low-energy fission.
\newblock Phys Rev C. 2017 Sep;96:034603.
\newblock Available from:
  \url{https://link.aps.org/doi/10.1103/PhysRevC.96.034603}.

\bibitem{Randrup2011a}
Randrup J, M\"oller P, Sierk AJ.
\newblock Fission-fragment mass distributions from strongly damped shape
  evolution.
\newblock Phys Rev C. 2011 Sep;84:034613.
\newblock Available from:
  \url{https://link.aps.org/doi/10.1103/PhysRevC.84.034613}.

\bibitem{Randrup2011b}
Randrup J, M\"oller P.
\newblock Brownian Shape Motion on Five-Dimensional Potential-Energy
  Surfaces:Nuclear Fission-Fragment Mass Distributions.
\newblock Phys Rev Lett. 2011 Mar;106:132503.
\newblock Available from:
  \url{https://link.aps.org/doi/10.1103/PhysRevLett.106.132503}.

\bibitem{Moller2015}
M\"oller P, Randrup J.
\newblock Calculated fission-fragment yield systematics in the region
  $74\ensuremath{\le}Z\ensuremath{\le}94$ and
  $90\ensuremath{\le}N\ensuremath{\le}150$.
\newblock Phys Rev C. 2015 Apr;91:044316.
\newblock Available from:
  \url{https://link.aps.org/doi/10.1103/PhysRevC.91.044316}.

\bibitem{Wahl1988}
Wahl AC.
\newblock Nuclear-charge distribution and delayed-neutron yields for
  thermal-neutron-induced fission of $^{235}$U, $^{233}$U, and $^{239}$Pu and
  for spontaneous fission of $^{252}$Cf.
\newblock Atomic Data and Nuclear Data Tables. 1988;39(1):1 -- 156.
\newblock Available from:
  \url{http://www.sciencedirect.com/science/article/pii/0092640X88900162}.

\bibitem{Ohsawa1999}
Ohsawa T, Horiguchi T, Hayashi H.
\newblock Multimodal analysis of prompt neutron spectra for $^{237}$Np(n,f).
\newblock Nuclear Physics A. 1999;653(1):17 -- 26.
\newblock Available from:
  \url{http://www.sciencedirect.com/science/article/pii/S0375947499001566}.

\bibitem{Ohsawa2000}
Ohsawa T, Horiguchi T, Mitsuhashi M.
\newblock Multimodal analysis of prompt neutron spectra for $^{238}$Pu(sf),
  $^{240}$Pu(sf), $^{242}$Pu(sf) and $^{239}$Pu(n$_{\rm th}$,f).
\newblock Nuclear Physics A. 2000;665(1):3 -- 12.
\newblock Available from:
  \url{http://www.sciencedirect.com/science/article/pii/S0375947499006867}.

\bibitem{Baba1997}
Baba H, Saito T, Takahashi N, Yokoyama A, Miyauchi T, Mori S, et~al.
\newblock Role of Effective Distance in the Fission Mechanism Study by the
  Double-energy Measurement for Uranium Isotopes.
\newblock Journal of Nuclear Science and Technology. 1997;34(9):871 -- 881.
\newblock Available from: \url{https://doi.org/10.1080/18811248.1997.9733759}.

\bibitem{Hambsch}
Hambsch FJ.
\newblock (personal communication);.

\bibitem{Kawano2010}
Kawano T, Talou P, Chadwick MB, Watanabe T.
\newblock Monte Carlo Simulation for Particle and $\gamma$-Ray Emissions in
  Statistical Hauser-Feshbach Model.
\newblock Journal of Nuclear Science and Technology. 2010 May;47(5):462 -- 469.
\newblock Available from:
  \url{http://www.tandfonline.com/doi/abs/10.1080/18811248.2010.9711637}.

\bibitem{Koning2003}
Koning AJ, Delaroche JP.
\newblock Local and global nucleon optical models from 1 keV to 200 MeV.
\newblock Nuclear Physics A. 2003;713(3):231 -- 310.
\newblock Available from:
  \url{http://www.sciencedirect.com/science/article/pii/S0375947402013210}.

\bibitem{Gilbert1965}
Gilbert A, Cameron AGW.
\newblock A composite nuclear-level density formula with shell corrections.
\newblock Can J Phys. 1965;43.
\newblock Available from:
  \url{http://www.nrcresearchpress.com/doi/abs/10.1139/p65-139#.V6IU247raQ8}.

\bibitem{Kawano2006}
Kawano T, Chiba S, Koura H.
\newblock Phenomenological Nuclear Level Densities using the KTUY05 Nuclear
  Mass Formula for Applications Off-Stability.
\newblock Journal of Nuclear Science and Technology. 2006;43(1):1 -- 8.
\newblock Available from:
  \url{http://www.tandfonline.com/doi/abs/10.1080/18811248.2006.9711062}.

\bibitem{RIPL3}
Capote R, Herman M, Oblo\v{z}insk\'{y} P, Young PG, Goriely S, Belgya T, et~al.
\newblock RIPL - Reference Input Parameter Library for Calculation of Nuclear
  Reactions and Nuclear Data Evaluations.
\newblock Nuclear Data Sheets. 2009;110(12):3107 -- 3214.
\newblock Available from:
  \url{http://www.sciencedirect.com/science/article/pii/S0090375209000994}.

\bibitem{Kopecky1990}
Kopecky J, Uhl M.
\newblock Test of gamma-ray strength functions in nuclear reaction model
  calculations.
\newblock Phys Rev C. 1990 May;41:1941 -- 1955.
\newblock Available from:
  \url{http://link.aps.org/doi/10.1103/PhysRevC.41.1941}.

\bibitem{INDC0603}
Herman M, Capote R, Sin M, Trkov A, Carlson BV, Oblo\v{z}insk\'{y} P, et~al.
\newblock EMPIRE-3.2 Malta, Modular system for nuclear reaction calculations
  and nuclear data evaluation User's Manual.
\newblock International Atomic Energy Agency; 2013. INDC(NDS)-0603.

\bibitem{Mumpower2017}
Mumpower MR, Kawano T, Ullmann JL, Krti\ifmmode~\check{c}\else \v{c}\fi{}ka M,
  Sprouse TM.
\newblock Estimation of $M1$ scissors mode strength for deformed nuclei in the
  medium- to heavy-mass region by statistical Hauser-Feshbach model
  calculations.
\newblock Phys Rev C. 2017 Aug;96:024612.
\newblock Available from:
  \url{https://link.aps.org/doi/10.1103/PhysRevC.96.024612}.

\bibitem{Pleasonton1972}
Pleasonton F, Ferguson RL, Schmitt HW.
\newblock Prompt Gamma Rays Emitted in the Thermal-Neutron-Induced Fission of
  $^{235}\mathrm{U}$.
\newblock Phys Rev C. 1972 Sep;6:1023 -- 1039.
\newblock Available from:
  \url{https://link.aps.org/doi/10.1103/PhysRevC.6.1023}.

\bibitem{Simon1990}
Simon G, Trochon J, Brisard F, Signarbieux C.
\newblock Pulse height defect in an ionization chamber investigated by cold
  fission measurements.
\newblock Nuclear Instruments and Methods in Physics Research Section A:
  Accelerators, Spectrometers, Detectors and Associated Equipment.
  1990;286(1):220 -- 229.
\newblock Available from:
  \url{http://www.sciencedirect.com/science/article/pii/016890029090224T}.

\bibitem{Straede1987}
Straede C, Budtz-J{\o}rgensen C, Knitter HH.
\newblock $^{235}$U(n,f) Fragment mass-, kinetic energy- and angular
  distributions for incident neutron energies between thermal and 6 MeV.
\newblock Nuclear Physics A. 1987;462(1):85 -- 108.
\newblock Available from:
  \url{http://www.sciencedirect.com/science/article/pii/0375947487903812}.

\bibitem{Zeynalov2006}
Zeynalov S, Furman W, Hambsch FJ.
\newblock Investigation of mass-TKE distributions of fission fragments from the
  U-235(n,f)- reaction in resonances.
\newblock ISINN-13. 2006;Available from:
  \url{http://isinn.jinr.ru/proceedings/isinn-13/pdf/351.pdf}.

\bibitem{Dyachenko1969}
D'yachenko PP, Kuzminov BD, Tarasko MZ.
\newblock Energy and mass distribution of fragments from fission of U-235 by
  monoenergetic neutrons from 0. to 15.5 MeV.
\newblock Soviet Journal of Nuclear Physics. 1969;8.

\bibitem{Wagemans1991}
Wagemans C.
\newblock The Nuclear Fission Process.
\newblock CRC Press; 1991.

\bibitem{Madland2006}
Madland DG.
\newblock Total prompt energy release in the neutron-induced fission of
  $^{235}$U, $^{238}$U, and $^{239}$Pu.
\newblock Nuclear Physics A. 2006;772(3):113 -- 137.
\newblock Available from:
  \url{http://www.sciencedirect.com/science/article/pii/S0375947406001503}.

\bibitem{Lestone2014}
Lestone JP, Strother TT.
\newblock Energy Dependence of Plutonium and Uranium Average Fragment Total
  Kinetic Energies.
\newblock Nuclear Data Sheets. 2014;118(Supplement C):208 -- 210.
\newblock Available from:
  \url{http://www.sciencedirect.com/science/article/pii/S0090375214000684}.

\bibitem{Duke2014}
Duke D.
\newblock Fision fragment mass distributions and total kinetic energy release
  of 235-uranium and 238-uranium in neutron-induced fission at intermediate and
  fast neutron energies. 2014;Ph.D Thesis, Colorado State University.

\bibitem{ANLNDM64}
Meadows JW, Budtz-J{\o}rgensen C.
\newblock The fission fragment angular distributions and total kinetic energies
  for 235-u(n,f) from .18 to 8.83 MeV.
\newblock Argonne National Laboratory; 1982. ANL/NDM-64.

\bibitem{Boldeman1967}
J~W~Boldeman AWD.
\newblock Prompt nubar measurements for thermal neutron fission.
\newblock AAEC-E. 1967;172.
\newblock Available from: \url{https://trove.nla.gov.au/work/35052889}.

\bibitem{Franklyn1978}
Franklyn CB, Hofmeyer C, Mingay DW.
\newblock Angular correlation of neutrons from thermal-neutron fission of
  $^{235}$U.
\newblock Physics Letters B. 1978;78(5):564 -- 567.
\newblock Available from:
  \url{http://www.sciencedirect.com/science/article/pii/0370269378906408}.

\bibitem{Diven1956}
Diven BC, Martin HC, Taschek RF, Terrell J.
\newblock Multiplicities of Fission Neutrons.
\newblock Phys Rev. 1956 Feb;101:1012 -- 1015.
\newblock Available from:
  \url{https://link.aps.org/doi/10.1103/PhysRev.101.1012}.

\bibitem{Holden1988}
Holden NE, Zucker MS.
\newblock Prompt Neutron Emission Multiplicity Distribution and Average Values
  (Nubar) at 2200 m/s for the Fissile Nuclides.
\newblock Nuclear Science and Engineering. 1988;98(2):174 -- 181.
\newblock Available from: \url{https://doi.org/10.13182/NSE88-A28498}.

\bibitem{Talou2011}
Talou P, Becker B, Kawano T, Chadwick MB, Danon Y.
\newblock Advanced Monte Carlo modeling of prompt fission neutrons for thermal
  and fast neutron-induced fission reactions on $^{239}\mathrm{Pu}$.
\newblock Phys Rev C. 2011 Jun;83:064612.
\newblock Available from:
  \url{http://link.aps.org/doi/10.1103/PhysRevC.83.064612}.

\bibitem{Rudstam1990}
Rudstam G, Aagaard P, Ekstr\"{o}m B, Lund E, G\"{o}kt\"{u}rk H, Zwicky HU.
\newblock Yields of Products from Thermal Neutron-Induced Fission of $^{235}$U.
\newblock Radiochimica Acta. 1990;49(4):155 -- 192.

\bibitem{Rudstam1987}
Rudstam G.
\newblock The determination of nuclear reaction yields by means of an
  isotope-separator-on-line (ISOL) system.
\newblock Nuclear Instruments and Methods in Physics Research Section A:
  Accelerators, Spectrometers, Detectors and Associated Equipment.
  1987;256(3):465 -- 483.
\newblock Available from:
  \url{http://www.sciencedirect.com/science/article/pii/0168900287902907}.

\bibitem{JENDL4}
Shibata K, Iwamoto O, Nakagawa T, Iwamoto N, Ichihara A, Kunieda S, et~al.
\newblock JENDL-4.0: A New Library for Nuclear Science and Engineering.
\newblock J Nucl Sci Technol. 2011 Jan;48:1 -- 30.
\newblock Available from:
  \url{http://www.tandfonline.com/doi/abs/10.1080/18811248.2011.9711675}.

\bibitem{Muller1984}
M\"uller R, Naqvi AA, K\"appeler F, Dickmann F.
\newblock Fragment velocities, energies, and masses from fast neutron induced
  fission of $^{235}\mathrm{U}$.
\newblock Phys Rev C. 1984 Mar;29:885 -- 905.
\newblock Available from:
  \url{https://link.aps.org/doi/10.1103/PhysRevC.29.885}.

\bibitem{Naik1995}
Naik H, Dange SP, Singh RJ, Datta T.
\newblock Systematics of fragment angular momentum in low-energy fission of
  actinides.
\newblock Nuclear Physics A. 1995;587(2):273 -- 290.
\newblock Available from:
  \url{http://www.sciencedirect.com/science/article/pii/0375947494008214}.

\bibitem{Ford1984}
Ford GP, Wolfsberg K, Erdal BR.
\newblock Independent yields of the isomers of $^{133}\mathrm{Xe}$ and
  $^{135}\mathrm{Xe}$ for neutron-induced fission of $^{233}\mathrm{U}$,
  $^{235}\mathrm{U}$, $^{238}\mathrm{U}$, and $^{242}\mathrm{Am}^{m}$.
\newblock Phys Rev C. 1984 Jul;30:195--213.
\newblock Available from:
  \url{https://link.aps.org/doi/10.1103/PhysRevC.30.195}.

\bibitem{Reeder1985}
Reeder PL, Warner RA, Ford GP, Willmes H.
\newblock Independent isomer yield ratio of $^{90}\mathrm{Rb}$ from thermal
  neutron fission of $^{235}\mathrm{U}$.
\newblock Phys Rev C. 1985 Oct;32:1327--1334.
\newblock Available from:
  \url{https://link.aps.org/doi/10.1103/PhysRevC.32.1327}.

\bibitem{Fogelberg1998}
Fogelberg B, Mach H, Gausemel H, Omtvedt JP, Mezilev KA.
\newblock New High Spin Isomers Obtained in Thermal Fission.
\newblock AIP Conference Proceedings. 1998;447:191.
\newblock Available from:
  \url{http://aip.scitation.org/doi/abs/10.1063/1.56702}.

\end{thebibliography}
\end{document}